# Phonon promoted charge density wave in topological kagome metal ScV$_6$Sn$_6$

Yong Hu[1,#,*], Junzhang Ma[2,3,4,#], Yinxiang Li[5,#], Dariusz Jakub Gawryluk[6], Tianchen Hu[7], Jérémie Teyssier[8], Volodymyr Multian[8], Zhouyi Yin[9], Yuxiao Jiang[10], Shuxiang Xu[7], Soohyeon Shin[6], Igor Plokhikh[6], Xinloong Han[11], Nicholas Clark Plumb[1], Yang Liu[12], Jiaxin Yin[13], Zurab Guguchia[14], Yue Zhao[9], Andreas P. Schnyder[15], Xianxin Wu[16,*], Ekaterina Pomjakushina[6], M. Zahid Hasan[10], Nanlin Wang[7,17,18], and Ming Shi[1,12,*]

[1]*Photon Science Division, Paul Scherrer Institut, CH-5232 Villigen PSI, Switzerland*
[2]*Department of Physics, City University of Hong Kong, Kowloon, Hong Kong, China*
[3]*City University of Hong Kong Shenzhen Research Institute, Shenzhen, China*
[4]*Hong Kong Institute for Advanced Study, City University of Hong Kong, Kowloon, Hong Kong, China*
[5]*College of Science, University of Shanghai for Science and Technology, Shanghai, 200093, China*
[6]*Laboratory for Multiscale Materials Experiments, Paul Scherrer Institut, CH-5232 Villigen PSI, Switzerland*
[7]*International Center for Quantum Materials, School of Physics, Peking University, Beijing 100871, China*
[8]*Department of Quantum Matter Physics, University of Geneva, 24 Quai Ernest-Ansermet, 1211 Geneva 4, Switzerland*
[9]*Institute for Quantum Science and Engineering and Department of Physics, Southern University of Science and Technology of China, Shenzhen, Guangdong 518055, China*
[10]*Laboratory for Topological Quantum Matter and Advanced Spectroscopy (B7), Department of Physics, Princeton University, Princeton, NJ, USA*
[11]*Kavli Institute for Theoretical Sciences, University of Chinese Academy of Sciences, Beijing 100190, China*
[12]*Center for Correlated Matter and Department of Physics, Zhejiang University, Hangzhou 310058, China*
[13]*Department of physics, Southern University of Science and Technology, Shenzhen, Guangdong 518055, China*
[14]*Laboratory for Muon Spin Spectroscopy, Paul Scherrer Institute, CH-5232 Villigen PSI, Switzerland*
[15]*Max-Planck-Institut für Festkörperforschung, Heisenbergstrasse 1, D-70569 Stuttgart, Germany*
[16]*CAS Key Laboratory of Theoretical Physics, Institute of Theoretical Physics, Chinese Academy of Sciences, Beijing 100190, China*
[17]*Beijing Academy of Quantum Information Sciences, Beijing 100913, China*
[18]*Collaborative Innovation Center of Quantum Matter, Beijing 100871, China*

#These authors contributed equally to this work.
*To whom correspondence should be addressed:
Y.H. (yonghphysics@gmail.com); X.W. (xxwu@itp.ac.cn); M.S. (ming.shi@psi.ch)

**Charge density wave (CDW) orders in vanadium-based kagome metals have recently received tremendous attention due to their unique properties and intricate interplay with exotic correlated phenomena, topological and symmetry-breaking states. However, the origin of the CDW order remains a topic of debate. The discovery of ScV$_6$Sn$_6$, a vanadium-based bilayer kagome metal exhibiting an in-plane $\sqrt{3}$ x $\sqrt{3}$ *R*30° CDW order with time-reversal symmetry breaking, provides a novel platform to explore the underlying mechanism behind the unconventional CDW. Here, we combine high-resolution angle-resolved photoemission spectroscopy, Raman scattering measurements and density functional theory to investigate the electronic structures and phonon modes of ScV$_6$Sn$_6$ and their evolution with temperature. We identify topologically nontrivial Dirac**

surface states and multiple van Hove singularities (VHSs) in the vicinity of the Fermi level, with one VHS near the $\overline{K}$ point exhibiting nesting wave vectors in proximity to the $\sqrt{3} \times \sqrt{3}$ $R30°$ CDW wave vector. Additionally, Raman measurements indicate a strong intrinsic electron-phonon coupling in ScV$_6$Sn$_6$, as evidenced by the presence of a two-phonon mode and a large frequency amplitude mode. Our findings highlight the fundamental role of lattice degrees of freedom in promoting the CDW in ScV$_6$Sn$_6$ and provide important insights into the fascinating correlation phenomena observed in kagome metals.

The kagome lattice, a corner-sharing triangle network, has emerged as a versatile platform for exploring unconventional correlated and topological quantum states. Due to the unique correlation effects and frustrated lattice geometry inherent to kagome lattices, several families of kagome metals have been found to display a variety of unconventional electronic instabilities and nontrivial topologies, including quantum spin liquid [1-3], unconventional superconductivity [4-7], charge density wave (CDW) orders [5-7], and Dirac/Weyl semimetals [8-10]. Of particular interest are the recently discovered non-magnetic vanadium-based superconductors $A$V$_3$Sb$_5$ ($A$=K, Rb, Cs), which exhibit intriguing similarities to correlated electronic phenomena observed in high-temperature superconductors, such as CDW [11-13], pair density wave [14], and electronic nematicity [15]. Especially, the three-dimensional (3D) CDW order with an in-plane 2×2 reconstruction possesses exotic properties, including time-reversal symmetry breaking [12,16,17], intertwined with unconventional superconductivity [17], and rotational symmetry breaking [13,14]. Two possible scenarios, namely phonon softening [18,19] and correlation-driven Fermi surface (FS) instability [20-23], have been proposed to account for the CDW order. However, despite intense research efforts, the origin of the CDW order and its symmetry-breaking characteristics remain elusive.

Very recently, a new family of vanadium-based bilayer kagome metals, $R$V$_6$Sn$_6$ (where $R$ represents a rare-earth element), has been discovered [24,25]. Although its kagome layer does not show long-range magnetic order, similar to $A$V$_3$Sb$_5$, magnetism can be introduced by controlling the $R$ sites, providing a tunable platform to investigate magnetism, nontrivial topology and correlation effects native to the kagome lattice. Notably, ScV$_6$Sn$_6$, a member of the bilayer kagome family, undergoes an intriguing 3D CDW phase transition with a wave-vector $\boldsymbol{Q} = (1/3, 1/3, 1/3)$ below $T_{CDW}$~92 $K$ [26], unlike the CDW observed in $A$V$_3$Sb$_5$. Interestingly, recent experimental evidence shows that time-reversal symmetry breaking also occurs in the CDW state [27]. However, the nature of the CDW and its driving force remain unresolved. An in-depth investigation of the band structure and its interplay with lattice vibrations in ScV$_6$Sn$_6$ would provide valuable insights into the mechanism underlying the CDW order with intriguing symmetry-breaking in kagome metals.

Here, we investigate the electronic and lattice degrees of freedom in the CDW formation of the kagome metal ScV$_6$Sn$_6$ using a combination of scanning tunneling microscopy (STM), high-resolution angle-resolved photoemission spectroscopy (ARPES), Raman scattering measurements and density functional theory (DFT). Our low-temperature STM topographs visualize an in-plane $\sqrt{3} \times \sqrt{3}$ $R30°$ reconstruction, corresponding to the bulk CDW wavevector measured in diffraction experiments [26]. In the electronic structure, we identify topologically nontrivial Dirac surface states (TDSSs) and multiple van Hove singularities (VHSs) in the vicinity of the Fermi level ($E_F$). Intriguingly, the nesting vector connecting the VHSs near the $\bar{K}$ point is close to (1/3, 1/3), matching with the observed $\sqrt{3}$ x $\sqrt{3}$ $R$30° CDW wave vector. In contrast to $A$V$_3$Sb$_5$, however, pronounced band reconstructions appear to be absent in the CDW state of ScV$_6$Sn$_6$, possibly due to the 3D nature of the $\sqrt{3}$ x $\sqrt{3}$ × 3 CDW order and a noticeable dispersion along the *c*-direction. Remarkably, our Raman measurements reveal the presence of a two-phonon mode in the normal state and Raman-active amplitude modes in the CDW phase, indicating a strong electron-phonon coupling. Collectively, our results emphasize the crucial role of lattice degrees of freedom in promoting the CDW in ScV$_6$Sn$_6$ and contribute to a deeper understanding of the diverse quantum correlation phenomena observed in vanadium-based kagome metals.

The pristine phase of ScV$_6$Sn$_6$ crystallizes in a layered structure with the space group P6/mmm. The unit cell consists of two V$_3$Sn[1] kagome layers, with Sn[2] and ScSn[3]$_2$ layers stacked in an alternating fashion along the out-of-plane direction (*c*-axis) (Fig. 1a). Similar to the sister compound GdV$_6$Sn$_6$, ScV$_6$Sn$_6$ tends to cleave along the *c*-axis, resulting in three surface terminations, namely the kagome, ScSn[3]$_2$ and Sn[2] layers [28,29]. The electrical resistivity [Fig. 1b(i)] and specific heat capacity [Fig. 1b(ii)] measurements consistently show a transition around 92 *K*, indicating the presence of the CDW transition [26,30]. To study the superlattice modulation of the CDW, we perform comparative STM measurements on ScV$_6$Sn$_6$ in both the normal state (Fig. 1c) and the CDW phase (Fig. 1d). Atomically resolved STM topographies clearly identify the hexagonal lattice formed by the Sn[2] atoms [Figs. 1c(i) and 1d(i)] and the in-plane modulation in the CDW phase [inset of Fig. 1d(i)]. The Fourier transform of the topographic data further visualizes the existence of a $\sqrt{3}$ x $\sqrt{3}$ $R$30° reconstruction in the CDW phase [Fig. 1d(ii)], which is absent in the normal state [Fig. 1c(ii)]. Figure 1e displays the bulk Brillouin zone (BZ) and the projected two-dimensional (2D) surface BZ. In the CDW phase, the in-plane component of the CDW folds the pristine BZ [Fig. 1f(i)] into the new smaller $\sqrt{3}$ x $\sqrt{3}$ BZ [Fig. 1f(ii)].

We next focus on the electronic structure of the bilayer kagome metal ScV$_6$Sn$_6$ (Fig. 2). Utilizing high-resolution ARPES measurements with a small beam spot, we reveal three different sets of ARPES spectra associated with the three possible surface terminations on the cleaved sample surface (Figs. 2a,2g, and Fig. S1). As previously established in GdV$_6$Sn$_6$ compound [29], the surface terminations of the sample can be identified by measuring the Sn 4*d* core level (Figs. 2b and 2h). In Figs. 2b-2f and Figs. 2h-2l, we present the electronic structure from the kagome (Fig. 2a) and ScSn[3]$_2$ terminations (Fig. 2g), respectively. The measured FS, with pristine BZ (the dashed red lines in Figs. 2c and 2i) and high-symmetry points ($\bar{\Gamma}$, $\bar{K}$ and $\bar{M}$) labeled, features a characteristic hexagonal kagome Fermiology, as generally exhibited in other kagome systems [8,10,11,29,31]. To visualize the energy-momentum dispersion of the electronics structures, polarization-dependent measurements were performed along two different high-symmetry paths, $\bar{\Gamma}$ - $\bar{M}$ (Figs. 2d and 2j) and $\bar{\Gamma}$ - $\bar{K}$ (Figs. 2e and 2k) directions.

Similar to other vanadium-based kagome metals [31,32], the photoemission intensities are strongly sensitive to the photon polarization (Figs. 2d-2e and 2j-2k), reflecting the multi-orbital nature of V-$d$ orbitals. In contrast to $A$V$_3$Sb$_5$, the photon-energy dependent measurements along the $\bar{\Gamma}$ - $\bar{K}$ direction (Figs. 2f and 2l) exhibits distinct band dispersions at different $k_z$ planes (also see Fig. S2), indicating the relatively obvious three-dimensionality of the electronic structure in ScV$_6$Sn$_6$.

A comparative examination of the band structure on different terminations reveals some differences, which can be attributed to the presence of surface states and matrix element effects. Specifically, on the ScSn$^3_2$ termination, the measured FS centered around the $\bar{\Gamma}$ point consists of a circular-shaped and a hexagonal-shaped FS sheet, as illustrated in Fig. 3a. A closer inspection of the band dispersion, as depicted in Fig. 3b, indicates that these FS sheets around $\bar{\Gamma}$ arise from two V-shaped bands, which bear a striking resemblance to the TDSSs previously observed in GdV$_6$Sn$_6$ [29]. Our DFT calculations (see Fig. S3 for details) confirm these observations and highlight that the observed V-shaped bands around the $\bar{\Gamma}$ point (Fig. 3b) are indicative of the existence of TDSSs originating from a $\mathbb{Z}_2$ bulk topology in ScV$_6$Sn$_6$.

Furthermore, in addition to the TDSSs, the ARPES spectra collected on the ScSn$^3_2$ termination reveal more details of the kagome bands, such as the characteristic Dirac cone (DC) and VHSs expected from the kagome tight-binding model [5,6]. Constant energy maps shown in Fig. 3c reveal two Dirac cones around the $\bar{K}$ point. The energy–momentum dispersion along the $\bar{\Gamma}$ - $\bar{K}$ - $\bar{M}$ - $\bar{\Gamma}$ direction (Fig. 3d), which agrees well with the calculated bulk states projected onto the (001) surface (Fig. 3e), confirms the existence of Dirac cones at binding energies ($E_B$) of 0.09 $eV$ (DC1) and 0.28 $eV$ (DC2), despite some differences in the energies of Dirac points between experimental data and theoretical calculations. Additionally, the band forming the DC2 extends to the $\bar{M}$ point and constitutes a VHS (labeled as VHS1 in Figs. 3d and 3e). The saddle point nature of VHS1 is evident from cuts taken vertically across the $\bar{K}$ - $\bar{M}$ path (#M$_1$-#M$_5$, as indicated in Fig. 3f), where the band bottom of the electron-like band (dashed green curve in Fig. 3g) exhibits a maximum energy slightly above $E_F$ at the $\bar{M}$ point (green solid curve). Furthermore, another hole-like band observed in Fig. 3h (same as Fig. 3g), which is slightly below the VHS1 band, has a minimum energy at the $\bar{M}$ point, indicating the electron-like nature along the orthogonal direction (the blue curve in Fig. 3h). This feature demonstrates another van Hove band with the opposite dispersion close to $E_F$ (marked as VHS2 in Figs. 3d). These twofold concavity VHSs are consistent with theoretical calculations (Fig. 3e) and have been identified in $A$V$_3$Sb$_5$ [31,33], where they are believed to promote the CDW order. Interestingly, we also identify an unusual VHS (referred to as VHS3) contributed by the DC1 band near the $\bar{K}$ point, as highlighted by the dashed red curve in Fig. 3d. To further confirm the saddle point nature of VHS3, we examine the band dispersions perpendicular to the $\bar{\Gamma}$ - $\bar{K}$ direction (cuts #K$_1$-#K$_5$ in Figs. 3f and 3i). The series of cuts in Fig. 3i reveal a hole-like band (dashed red curve) with a band bottom that exhibits a minimum energy at the $\bar{A}$ point. These are fully consistent with our calculations shown in the inset of Fig. 3e, confirming its van Hove nature (see also Fig. S4). Due to the six-fold rotational symmetry of the lattice, there are six such saddle points near $\overline{K}$ and $\overline{\Lambda}$, as shown in Fig. 3j(ii).

As VHSs carry large density of states and can promote competing electronic instabilities, we now explore the potential contribution of the identified multiple VHSs to the CDW in $ScV_6Sn_6$. Previous theoretical studies have emphasized that VHSs located at the $M$ point can naturally give rise to nesting vectors $Q_{1,2,3}$ [5,6] that connect different sublattices on the saddle points of the FS [Fig. 4a(i)], potentially leading to a 2 × 2 bond CDW instability. However, we note that the suggested FS nesting wave vectors $Q_{1,2,3}$ in Fig. 3j(i) are incompatible with the in-plane $\sqrt{3}$ x $\sqrt{3}$ $R30°$ reconstruction observed in $ScV_6Sn_6$ [Fig. 1d(ii)]. Nevertheless, the nesting vectors associated with the identified VHS3 near the $\overline{K}$ point are in proximity to (1/3,1/3) [Fig. 3j(ii)], which is more consistent with the observed in-plane $\sqrt{3}$ x $\sqrt{3}$ CDW pattern. To assess the role of VHS3 in the CDW formation, we perform temperature-dependent measurements on the band dispersions along the $\overline{\Gamma}$ - $\overline{K}$ direction. Surprisingly, our high-resolution ARPES spectra show negligible differences between the CDW phase and normal state (Figs. 3k and 3l), in contrast to the significant band reconstructions observed in $AV_3Sb_5$ [32,34]. As the V-3$d$ states dominate near the $E_F$, this weak band reconstruction and folding effect may be due to the 3D nature of the $\sqrt{3}$ x $\sqrt{3}$ × 3 CDW order, where the distortion of V atoms is small [26], and noticeable dispersion along the $c$-direction (Fig. S2).

After investigating the electronic structure of $ScV_6Sn_6$, we next examine the effects of the lattice degrees of freedom on the CDW formation using Raman scattering. Figure 4a displays a colormap of the Raman response, covering a temperature from 24 $K$ to 200 $K$. The Raman spectra (Figs. 4a and 4b) feature two prominent Raman-active phonon peaks at 143 cm$^{-1}$ and 243 cm$^{-1}$, which we attribute to the $E_{2g}$ and $A_{1g}$ modes, respectively, based on the polarization-dependent measurements (for details see Fig. S5) and theoretical calculations (Fig. 4c). Additionally, we observe multiple weak peak-like structures below 100 cm$^{-1}$ (Fig. S5) and a broad peak around 116 cm$^{-1}$ (highlighted by the dashed white curve in Fig. 4a and red arrow in Fig. 4b) in the spectra above $T_{CDW}$ (indicated by the dashed red line in Fig. 4a). The weak peaks below 100 cm$^{-1}$, showing almost no temperature dependence (Fig. S5), likely arise from rotational spectrum of $O_2$ and $N_2$ along the laser path [35]. As the temperature decreases, both the $E_{2g}$ and $A_{1g}$ modes exhibit a blueshift (Fig. S5), while the broad peak around 116 cm$^{-1}$ shifts minimally above $T_{CDW}$, but abruptly vanishes below $T_{CDW}$ (Fig. 4b). The temperature-dependent behavior of the broad peak resembles the one observed in other well-studied CDW materials [36-39], indicating the presence of a two-phonon Raman mode. This mode involves two phonons with opposite wave vectors and represents a second order process usually correlated with the strong momentum dependent electron-phonon coupling near the CDW wave vector [37,40-42]. In $ScV_6Sn_6$, the observed two-phonon mode likely originates from the acoustic longitudinal modes in the $K$-$H$ path [Fig. 4f(ii), the shaded region highlights the half frequency of the two-phonon mode], according to the theoretical phonon dispersion in Fig. 4f. Below $T_{CDW}$, the two-phonon mode disappears, possibly due to CDW-induced phonon folding and alteration of electron-phonon coupling.

Moreover, multiple new phonon peaks (labeled as $A_1$-$A_4$ in Fig. 4b) emerge below $T_{CDW}$, around 150 cm$^{-1}$ and 240 cm$^{-1}$, indicating their intimate relationship with the CDW order. Interestingly, the two new modes ($A_1$, $A_2$) close to the $E_{2g}$ have almost no specific temperature dependence in their frequencies and linewidths (Figs. 4a and 4b) as the temperature approaches $T_{CDW}$, consistent with characteristics of CDW zone-folded modes. In contrast, the $A_3$ mode shows noticeable softening and broadening with warming towards $T_{CDW}$ (Figs. 4d and 4e), eventually becoming unresolvable above $T_{CDW}$ (Figs. 4a, 4b, 4d and 4e). These are indicative of a CDW amplitude mode derived from the collapse of coherent CDW order near $T_{CDW}$ [36-39,43].

Our theoretical calculations show that imaginary phonon modes appear at the $H$ and $L$ points (Fig. 4f), corresponding to $\sqrt{3}$ x $\sqrt{3}$ x 2 and 2 x 2 x 2 lattice reconstructions, respectively. However, these modes, along with the absence of unstable phonon modes at (1/3, 1/3, 1/3) (Fig. 4f), fail to explain the observed $\sqrt{3}$ x $\sqrt{3}$ × 3 CDW order. This suggests that the bare phonon instability is insufficient to account for the CDW order in our experiments (Figs. 1b and 1d). Our identification of the two-phonon mode order, typically much weaker than the one-phonon Raman modes, points to a strong electron-phonon coupling in ScV$_6$Sn$_6$. This coupling could induce significant phonon softening at the CDW vector by introducing a negative self-energy term through the electron bubble. Consequently, the renormalized phonon dispersion may exhibit an anomaly and a minimum negative frequency at the wave vector $\boldsymbol{Q}$ = (1/3, 1/3, 1/3), giving rise to the observed CDW order. Our Raman measurements support this scenario, as they show the absence of one-phonon softening modes and the observation of amplitude modes with high frequency [44]. Furthermore, the nesting vector between the observed VHSs around the $\bar{K}$ point [Fig. 3j(ii)] aligns with the in-plane component of the CDW vector, suggesting that electronic correlation may also participate in promoting the in-plane component of CDW order [45].

Finally, we discuss the possible origin of time-reversal symmetry breaking charge order in the non-magnetic vanadium-based kagome metals $A$V$_3$Sb$_5$ [12,16,17] and ScV$_6$Sn$_6$ [27]. While the CDW order in both systems shares some similarities, there are two main differences. First, in $A$V$_3$Sb$_5$, the CDW pattern mainly results from the distortion of the kagome V atoms (Fig. 4g) [32,46,47]. In contrast, in ScV$_6$Sn$_6$, the CDW order mainly involves the displacement of Sc and Sn, while the V atoms show negligible distortion (Fig. 4h) [26]. Second, the presence of a two-phonon mode and a large frequency amplitude mode in ScV$_6$Sn$_6$ suggests a much stronger electron-phonon coupling than in $A$V$_3$Sb$_5$. These observations imply that electron-phonon coupling may play a crucial role in promoting the CDW order in ScV$_6$Sn$_6$. Further considering the correlation effect associated with VHSs, we deduce that electron-electron interactions and electron-phonon coupling may conspire to generate the symmetry-breaking states in the vanadium-based kagome metals, which warrants further investigations, both from the theoretical and the experimental fronts.

In conclusion, our study combining ARPES and Raman scattering measurements provides important insights into the underlying mechanism of the CDW order in ScV$_6$Sn$_6$. The VHSs located near the $\bar{K}$ point introduce nesting wave vectors close to (1/3,1/3), which are consistent with the observed in-plane $\sqrt{3} \times \sqrt{3}$ CDW order. Furthermore, our Raman measurements demonstrate the presence of the two-phonon mode and amplitude modes, suggesting a strong electron-phonon coupling [40]. Taken together, our results suggest a concerted mechanism of the CDW order in ScV$_6$Sn$_6$ involving both electron-phonon coupling and electron correlation effects. Further investigations are necessary to fully comprehend the interplay between these two mechanisms and their roles in promoting the unconventional CDW order in vanadium-based kagome metals.

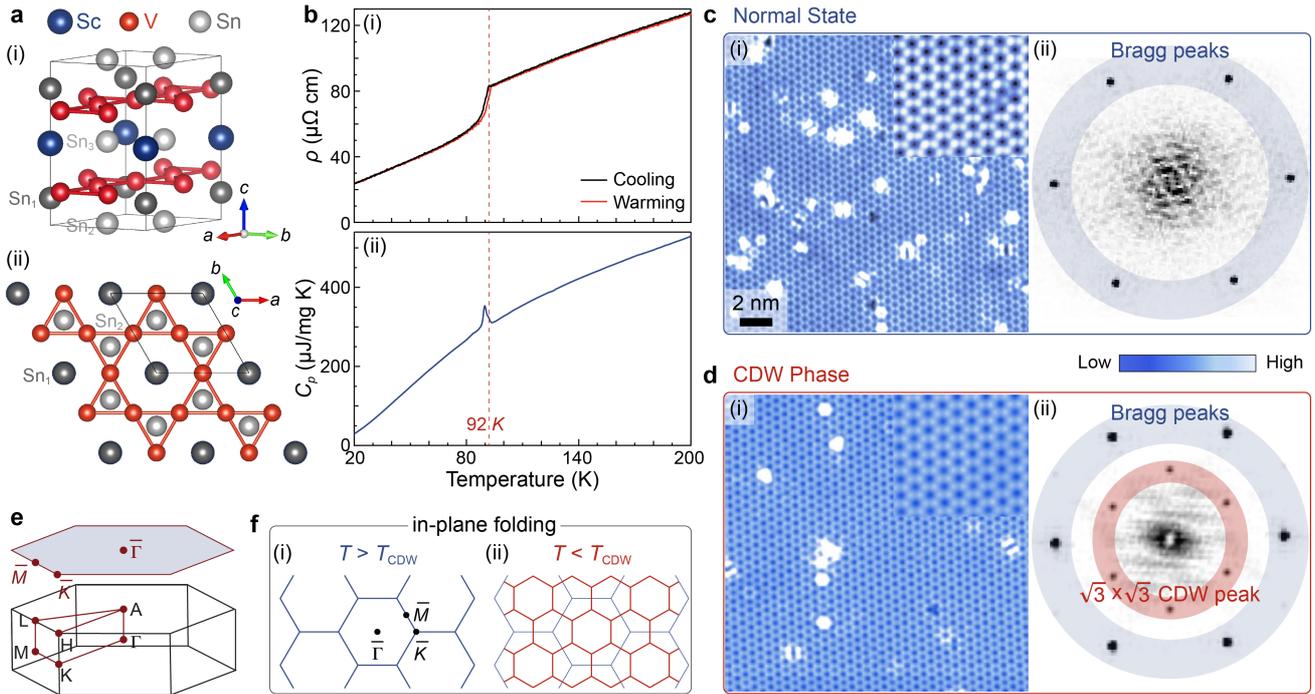

**Fig. 1. Crystal structure, transport and topographic characterizations of ScV₆Sn₆.** **a** Crystal structure in the normal state showing the unit cell (i) and top view displaying the kagome lattice (ii). **b** Temperature-dependent *ab*-plane resistivity (i) and specific heat capacity (ii) of ScV₆Sn₆, indicating the onset of CDW near 92 *K*. **c** STM topograph of Sn² termination measured at 80 *K* (i) and associated Fourier transforms (ii). Atomic Bragg peaks are highlighted with a blue ring. **d** Same as I, but taken at 1 *K*. √3 x √3 *R*30° CDW peaks are marked with a red ring. **e** Schematic of the bulk and surface Brillouin zones (BZs), with high-symmetry points marked. **f** Schematics of the in-plane folding of the surface BZ. Pristine (i) and CDW (ii) BZs are shown with blue and red lines, respectively.

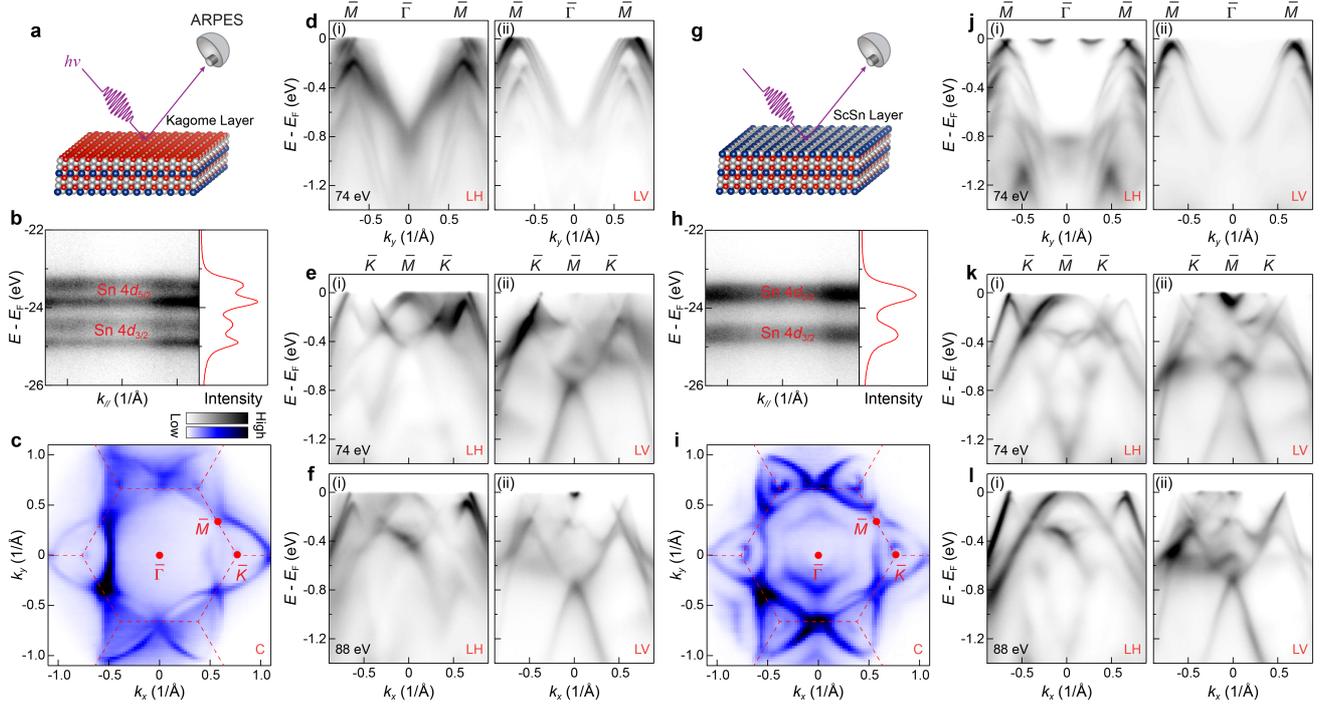

**Fig. 2. Termination-dependent photoemission measurements of the electronic structure in ScV$_6$Sn$_6$. a** Schematic of the kagome termination. **b** The corresponding x-ray photoelectron spectroscopy (XPS) spectrum on the Sn 4*d* core levels (left) and the integrated energy distribution curve (EDC) of the core levels (right). **c** Fermi surface (FS) mapping collected on the kagome termination, measured with circular (C) polarized light. The dashed red line represents the pristine BZ. **d** Photoelectron intensity plots of the band structure taken along the $\bar{\Gamma}$ - $\bar{M}$ direction on the kagome termination, measured with 74 *eV* linear horizontal (LH) (i) and linear vertical (LV) (ii) polarizations. **e,f** Same as (d), but taken along the $\bar{\Gamma}$- $\bar{K}$ direction, measured with 74 *eV* (e) and 88 *eV* (f). **g-l** Same as (a-f), but measured on the ScSn$^3_2$ termination.

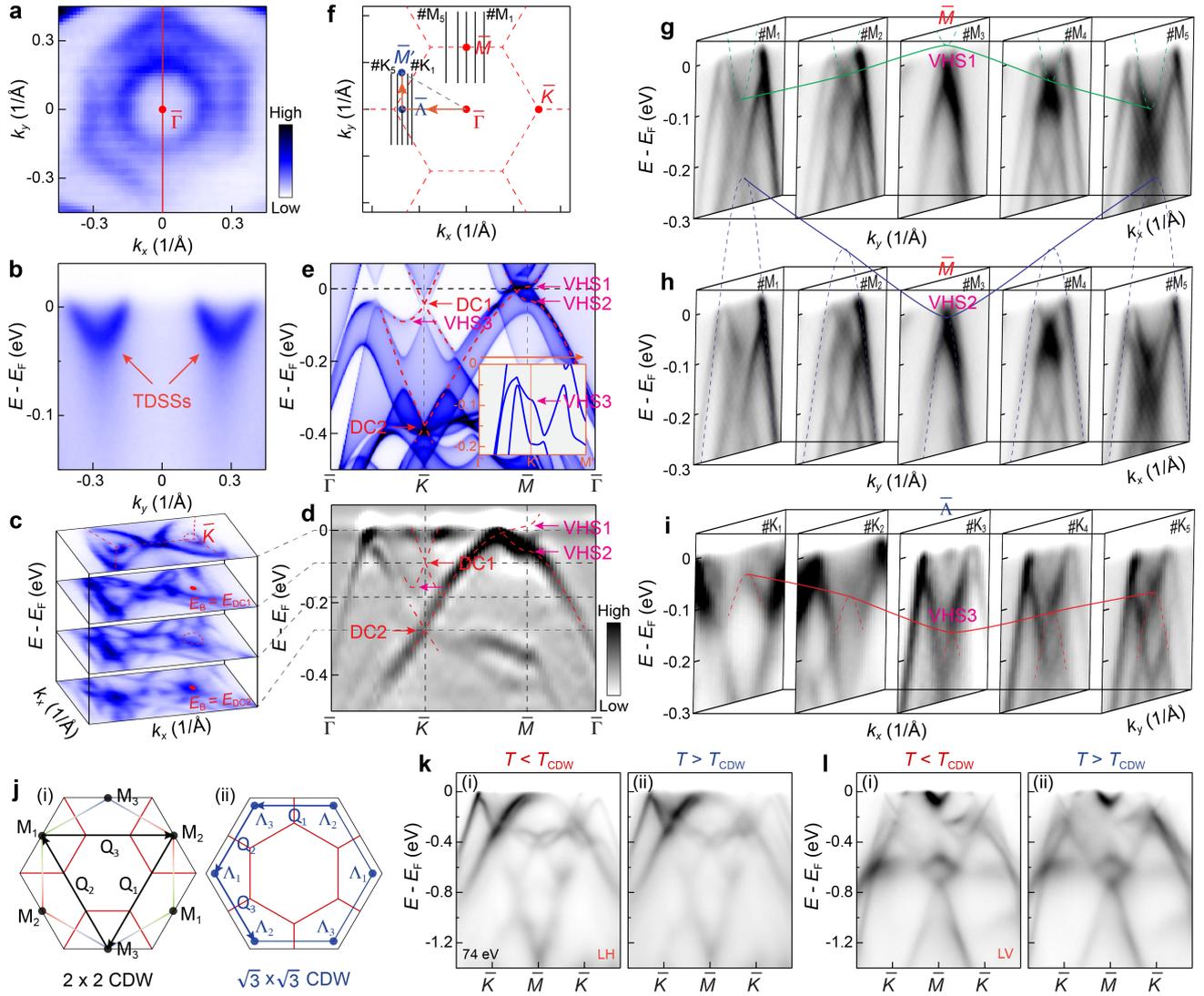

**Fig. 3.** $\mathbb{Z}_2$ **topological surfaces states, Dirac cones and VHSs in ScV$_6$Sn$_6$. a** Zoom-in FS mapping measured on the Sn termination. **b** ARPES spectrum taken along the $\bar{\Gamma}$ - $\bar{M}$ direction highlighting the TDSSs. The momentum path is indicated by the red line in (a). **c** Stacking plots of constant energy maps around the $\bar{K}$ point. **d** Energy–momentum dispersion along the $\bar{\Gamma}$ - $\bar{K}$ - $\bar{M}$ - $\bar{\Gamma}$ direction. **e** Calculated band structure along the $\bar{\Gamma}$ - $\bar{K}$ - $\bar{M}$ - $\bar{\Gamma}$ direction. Dirac cone (DC) and VHS are indicated by red and pink arrows, respectively. The inset displays DFT bands along the $\Gamma$ - $\Lambda$ - $M'$ direction [as indicated by the orange arrow in (f)]. **f** Schematics of the surface BZ. **g** A series of cuts taken vertically across the $\bar{K}$ - $\bar{M}$ path, the momentum paths of the cuts (#M$_1$-#M$_5$) are indicated by the black lines in (f). Dashed green curve highlights the electron-like band and solid curve indicates the corresponding VHS1 at the $\bar{M}$ point. **h** Same as (g), but highlights the hole-like band (dashed blue curve) and VHS2 (solid curve). **i** Stack of cuts perpendicular to the $\bar{\Gamma}$ - $\bar{K}$ direction. The momentum directions of the cuts (#K$_1$-#K$_5$) are indicated by the black lines in (f). Dashed red curve and solid curve indicate the hole-like band and corresponding VHS3 at the $\bar{\Lambda}$ point, respectively. The ARPES spectra shown in (g-i) were taken with 74 eV C polarized light. **j** FS of kagome lattice at the VHS filling. The three inequivalent saddle points M$_i$ ($\Lambda_i$) are connected by three inequivalent nesting vectors Q$_i$, which can give rise to a 2 x 2 CDW (i) and $\sqrt{3}$ x $\sqrt{3}$ CDW (ii). **k** Temperature-dependent measurements of the band structure along the $\bar{\Gamma}$ - $\bar{K}$ direction, measured below $T_{CDW}$ at 20 K (i) and above $T_{CDW}$ at 130 K (ii) with 74 eV LH polarized light. **l** Same as (b), but measured with LV polarization.

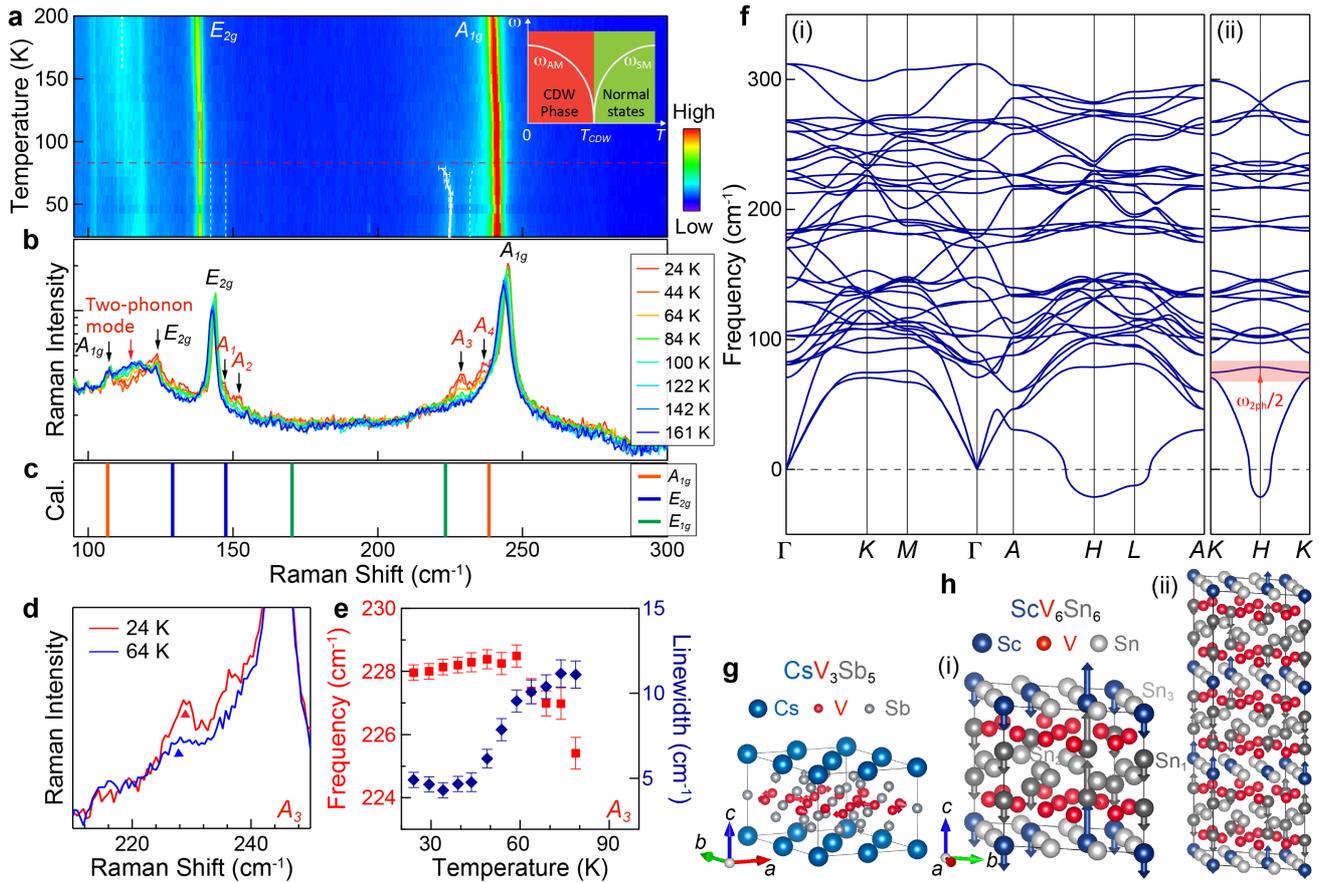

**Fig. 4. Raman modes and phonon band structure in ScV$_6$Sn$_6$. a** Temperature-dependent colormap of the Raman response recorded on ScV$_6$Sn$_6$. The inset illustrates the relationship between the soft mode and amplitude mode in typical CDW materials. The soft mode frequency ($\omega_{SM}$) freezes below $T_{CDW}$, and the amplitude mode frequency ($\omega_{AM}$) emerges afterward. **b** Typical Raman spectra measured below and above $T_{CDW}$. **c** Calculated Raman mode frequencies around Γ point. **d** Raman spectra measured at 24 K and 64 K. The triangle indicates the $A_3$ phonon peak. **e** Temperature dependence of the Frequency and linewidth of the $A_3$ mode. **f** DFT calculated phonon band structure along high-symmetry paths (i) and the K - H - K path (ii) of pristine ScV$_6$Sn$_6$, with experimental lattice parameters [26]. The red shaded region indicates the half frequency of the two-phonon mode. **g** Distortion pattern of the trihexagonal pattern in the 2 x 2 x 1 CDW phase of CsV$_3$Sb$_5$. **h** Acoustic phonon mode at the K point (i) corresponding to the observed two-phonon mode in (a, b) and distortion pattern (ii) of the $\sqrt{3} \times \sqrt{3} \times 3$ CDW indicated by the vectors, with respect to the pristine phase of ScV$_6$Sn$_6$ [Fig. 1a(i)]. The length of the vectors represents the amplitude of atomic displacements.